\documentclass[aps,showpacs,twocolumn]{revtex4-2}
\usepackage{amsmath}
\usepackage{amssymb}
\usepackage[dvipsnames]{xcolor}
\usepackage[many]{tcolorbox}
\usepackage{array}
\usepackage{multirow}
\usepackage{makecell}
\usepackage{hyperref}
\usepackage{booktabs}
\usepackage{array}

\hypersetup{hypertex=true, colorlinks=true, linkcolor=blue, urlcolor=blue, citecolor=blue}

\begin{document}

\title{Pairing-induced Bloch oscillations in an interacting Kitaev chain}
\author{E. S. Ma and Z. Song}
\email{songtc@nankai.edu.cn}
\affiliation{School of Physics, Nankai University, Tianjin 300071, China}

\begin{abstract}
We study the peculiar dynamics of the Kitaev chain induced by
nearest-neighbor (NN) interaction. We show that a strong NN interaction
suppresses single-particle hopping but enhances pairing, resulting in a
Wannier-Stark ladder. Based on the spin-fermion correspondence at the
symmetry point, the model maps to a transverse field Ising model on a zigzag
lattice, providing a clear physical picture and guiding experimental
verification. The Wannier-Stark state corresponds to a localized domain wall
between ferromagnetic and antiferromagnetic phases. It exhibits Bloch
oscillation even in the absence of a longitudinal field, in contrast to
previous works. Numerical simulations of time-dependent observables verify
these conclusions. Our findings provide an example demonstrating emergent
Stark many-body localization.
\end{abstract}

\maketitle

\section{Introduction}

\label{Introduction}

The Wannier-Stark ladder (WSL) represents a cornerstone of quantum transport
theory, connecting fundamental quantum mechanics to modern device physics
and serving as a paradigmatic example of how periodic potentials and
external fields combine to produce unexpected quantum behavior \cite%
{Bloch1929,wannier1959elements,Wannier1960,Glueck2002,Waschke1993}. It
provides the theoretical basis for Bloch oscillations, the remarkable
phenomenon where electrons in a perfect crystal under a DC electric field
exhibit an AC response with oscillations at the Bloch frequency. This
counterintuitive quantum effect has been observed in systems of
semiconductor superlattices and ultracold atoms \cite%
{Waschke1993,BenDahan1996,Wilkinson1996,Anderson1998}. In the past decade,
such a phenomenon has attracted much attention in cold-atom physics and
photonics due to applications in interferometric measurements and as a
method for manipulating localized wave packets \cite%
{Breid2006,Breid2007,Dreisow2009,Kling2010,Ploetz2011}. It can be simulated
using artificial quantum systems, such as superconducting circuits \cite%
{Song2024}. Meanwhile, the dynamics of particle pairs in lattice systems
have attracted significant attention, driven by rapid progress in
experimental techniques. Ultra-cold atoms serve as an exceptional platform
for exploring few-particle physics, since optical lattices enable clean
implementations of diverse many-body Hamiltonians. It stimulates many
experimental \cite{Winkler2006,Foelling2007,Gustavsson2008} and theoretical
investigations \cite%
{Mahajan2006,Petrosyan2007,Creffield2007,Kuklov2007,Zoellner2008,Wang2008,Valiente2008,Jin2009,Valiente2009,Valiente2010,Javanainen2010,Wang2010,Rosch2008,Zhang2024}
in strongly correlated systems. It has recently been established that
subjecting a bound pair to a linear potential induces periodic dynamics,
even within a correlated system \cite%
{Khomeriki2010,Longhi2011,Longhi2012,Corrielli2013,Lin2014,ZhangXZ2016,Zhang2024}%
. In addition, a theorem establishing the existence of WSL for a broad class
of systems, including strongly correlated and non-Hermitian systems with
conserved particle number, has been rigorously proven in quantum many-body
settings \cite{Zhang2025Bloch,Zhang2015formation,Zhang2024extended}.
However, whether WSL exists in systems without particle number conservation
remains an open question. This question may be addressed by examining WSL in
the transverse field Ising chain, which is equivalent to a spinless-fermion
Kitaev chain where particle number is not conserved. Nevertheless, the
existence of WSL in the proposed systems requires a longitudinal field that
takes an unphysical form under the Jordan-Wigner transformation \cite%
{Zhang2025Bloch,Zhang2015formation,Zhang2024extended}.

In this work, we extend the concept of WSL to many-body systems without
particle number conservation. We investigate the formation of WSL in an
interacting Kitaev chain. We show that such a system shares the same
spectrum as a transverse field Ising chain with next-nearest-neighbor (NNN)
coupling, based on the domain-wall representation \cite{ma2026boundary}. The
distinction of this model from previous work is that the WSL can be formed
without the longitudinal field. In the spinless-fermion representation, the
underlying mechanism of WSL is the nearest-neighbor (NN) pairing process
under resonant conditions. On the other hand, in the quantum spin
representation, it corresponds to the dynamics of the domain wall between
the N\'{e}el state and ferromagnetic state. Numerical simulations of the
time evolution of typical initial states validate our theoretical
predictions. The conclusion can be applied to the case with multiple domain
walls. The NN interaction plays the dominant role in preventing particles
from delocalizing and spreading across the system. It causes localization
without the need for external random disorder and a strong tilted potential,
thus offering an example of emergent Wannier-Stark many-body localization.

This paper is organized as follows. In Sec. \ref{Model and Wannier-Stark
ladder}, we introduce an interacting Kitaev model and demonstrate that the
Hamiltonian reduces to a Wannier-Stark ladder with equal energy spacing
within a subspace under a certain parameter setting. In Sec. \ref{Quantum
spin representation}, we establish a map between the interacting Kitaev
model and a spin chain with NNN interactions and a transverse field by
interpreting a magnetic domain wall as a particle excitation, and we review
the exact solution obtained in previous work from an alternative
perspective. Sec. \ref{Bloch oscillations} is dedicated to the discussions
of periodic dynamics for two kind of initial states and their numerical
verification. Finally, we provide a summary in Sec. \ref{Summary}.

\section{Model and Wannier-Stark ladder}

\label{Model and Wannier-Stark ladder}

We consider an interacting spinless fermion Kitaev model on an $N$-site
chain. The Hamiltonian consists of two parts 
\begin{equation}
H=H_{\mathrm{V}}+H_{\mathrm{T}},  \label{H}
\end{equation}%
where $H_{\mathrm{T}}$\ describes the hopping and pairing terms 
\begin{equation}
H_{\mathrm{T}}=\sum_{j=1}^{N-1}\left( \tau c_{j}^{\dagger }c_{j+1}+\Delta
c_{j}^{\dagger }c_{j+1}^{\dagger }\right) +\mathrm{H.c.},  \label{HT}
\end{equation}%
%
%
%
%
%
and $H_{\mathrm{V}}$\ contains the interacting and on-site potential terms%
\begin{equation}
H_{\mathrm{V}}=\sum_{j=1}^{N-1}U\left( 2n_{j}-1\right) \left(
2n_{j+1}-1\right) -\mu \sum_{j=1}^{N}\left( n_{j}-{\frac{1}{2}}\right) .
\end{equation}%
In the particle-number basis, $H_{\mathrm{V}}$\ and $H_{\mathrm{T}}$\
correspond to the diagonal and off-diagonal elements of the matrix
representation of the Hamiltonian. Here $c_{j}^{\dagger }$ and $c_{j}$ are
the fermionic creation and annihilation operators at site $j$, $%
n_{j}=c_{j}^{\dagger }c_{j}$ is the fermion occupation number operator, $\tau
$ is the hopping integral, $\Delta $ is the $p$-wave superconducting pairing
strength, $\mu $ is the chemical potential, and $U$ is the nearest-neighbor
density-density interaction strength.

At $U=0$, the model reduces to the usual non-interacting Kitaev chain model 
\cite{Kitaev2001}, which can be exactly diagonalized by means of the
Majorana transformation. It is a theoretical model that describes a
one-dimensional chain of superconducting nanowires with $p$-wave pairing
symmetry. It is one of the key models in condensed matter physics,
particularly for studying topological superconductivity and Majorana
fermions. The exact solution shows that the system has a phase transition
between a topologically trivial phase and a topologically non-trivial phase.
The transition is typically controlled by the parameters $\tau$, $\Delta $,
and $\mu $. For the interacting case with $U\neq 0$, no exact solution is
generally available. It has been shown that the ground states can be
constructed when the chemical potential $%
\mu
$ takes a specific value determined by the other parameters ($\tau$, $\Delta $,
and $U$) \cite{Katsura2015}. In addition, the model is exactly solvable at
the symmetry point $\Delta = \tau $ with $\mu =0$ under open boundary
condition \cite{Miao2017}.

In the following, we focus on the dynamical properties of this system from
another perspective. We start with the Hamiltonian $H_{\mathrm{V}}$, which
is diagonal in the particle-number basis. Any state of the form $%
c_{i}^{\dagger }c_{j}^{\dagger }...c_{k}^{\dagger }\left\vert 0\right\rangle 
$, a product of creation operators acting on the vacuum state $\left\vert
0\right\rangle $, is an eigenstate of $H_{\mathrm{V}}$. Of particular
interest is a set of eigenstates $\left\{ \left\vert \phi _{l}\right\rangle
:l\in \left[ 1,N\right] \right\} $ with corresponding energy $E_{l}$, where%
\begin{equation}
\left\vert \phi _{l}\right\rangle =\prod_{j=1}^{l}c_{j}^{\dagger }\left\vert
0\right\rangle ,
\end{equation}%
and the energies are given by 
\begin{equation*}
E_{l}=\left( N-2l\right) \mu /2+\left( N-3\right) U 
\end{equation*}%
for $l\in \left[ 1,N\right) .$ To provide an intuitive picture, these states
can also be represented as 
\begin{equation*}
\left\vert \phi _{l}\right\rangle =|\underset{l}{\underbrace{\bullet \bullet
...\bullet \bullet }}\circ \circ ...\circ \circ \rangle ,
\end{equation*}%
where $\circ $\ and $\bullet $\ denote empty and filled sites, respectively.

When the term $H_{\mathrm{T}}$\ is turned on, we have 
\begin{equation}
H_{\mathrm{T}}\left\vert \phi _{2j}\right\rangle =\Delta \left\vert \phi
_{2j-2}\right\rangle +\Delta \left\vert \phi _{2j+2}\right\rangle ,
\end{equation}%
for $j\in \left[ 2,N/2-2\right] $ and 
\begin{equation}
H_{\mathrm{T}}\left\vert \phi _{2j-1}\right\rangle =\Delta \left\vert \phi
_{2j-3}\right\rangle +\Delta \left\vert \phi _{2j+1}\right\rangle ,
\end{equation}%
for $j\in \left[ 2,N/2-1\right] $, describing transitions between energy
levels $E_{l}$ and $E_{l\pm 2}$. The system thus possesses two invariant
subspaces: one spanned by states $\left\vert \phi _{l}\right\rangle $, with
even $l$, and the other by states with odd $l$. The corresponding effective
Hamiltonian takes the same form in these two subspaces:%
\begin{eqnarray}
H_{\mathrm{eff}} &=&\Delta \sum_{j=1}^{N/2-2\left( N/2-1\right) }\left(
\left\vert j+1\right\rangle \left\langle j\right\vert +\mathrm{H.c.}\right) 
\notag \\
&&-2\mu \sum_{j=1}^{N/2-1\left( N/2\right) }j\left\vert j\right\rangle
\left\langle j\right\vert ,  \label{H_eff}
\end{eqnarray}%
where the basis states are defined as $\left\vert j\right\rangle =\left\vert
\phi _{2j}\right\rangle $ for the even subspace and $\left\vert
j\right\rangle =\left\vert \phi _{2j-1}\right\rangle $ for the odd subspace.
A constant energy shift has been neglected: $N\mu /2+\left( N-3\right) U$
for even $l$, and $\left( N+2\right) \mu /2+\left( N-3\right) U$ for odd $l$%
. This is a standard tight-binding Hamiltonian describing a Wannier-Stark
ladder system in the large $N$ limit.

It is well known that the spectrum of $H_{\mathrm{eff}}$ is equally spaced
levels, which supports periodic dynamics \cite{Hartmann2004,Bloch1929,1934}.
Such Wannier-Stark ladders have been realized in semiconductor superlattices
and ultracold atoms \cite%
{Waschke1993,BenDahan1996,Wilkinson1996,Anderson1998}. The eigenstates $%
\left\vert \psi _{n}\right\rangle $\ satisfying the Schr\"{o}dinger equation 
\begin{equation}
H_{\mathrm{eff}}\left\vert \psi _{n}\right\rangle =E_{n}\left\vert \psi
_{n}\right\rangle ,
\end{equation}%
are explicitly expressed in terms of the Bessel function $J_{m}$\ as%
\begin{equation}
\left\vert \psi _{n}\right\rangle =\sum_{m}J_{m-n}(\frac{\Delta }{\mu }%
)|m\rangle ,  \label{eigenstate}
\end{equation}%
with $E_{n}=-2n\mu $. We note that the eigen energy $E_{n}$ is independent
of $\Delta $, and shifts to lower energies when $U$ is negative.

\begin{figure}[tbh]
\centering
\includegraphics[width=0.5\textwidth]{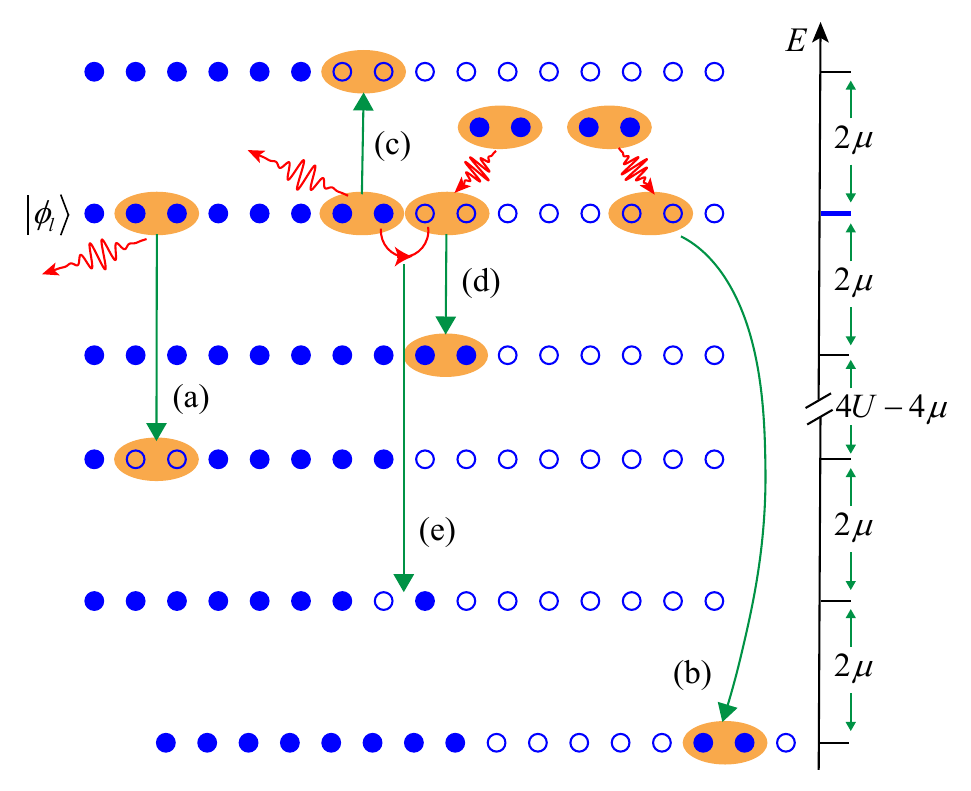}
\caption{The illustration of the effectiveness of $H_{\mathrm{eff}}$ in the
subspace $\left\{ \left\vert \protect\phi _{l}\right\rangle \right\}$. For a
given basis $\left\vert \protect\phi _{l}\right\rangle$, where $l$ sites on
the left of the chain are occupied by fermions (solid blue circles) and the
rest of the sites are empty (hollow blue circles), there are five cases when
the Hamiltonian $H_{\mathrm{T}}$ in Eq. (\protect\ref{HT}) acts on it. The
right energy axis marks the corresponding energy change. (a) and (b) denote
the pairing-elimination and pairing-creation processes that transform this
state into another subspace, with the corresponding energy change under $H_{%
\mathrm{V}}$ being $-4U\pm2\protect\mu$; (c) and (d) represent the
pairing-elimination and pairing-creation that preserve the subspace and
shift the energy by $\pm2\protect\mu$; (e) shows the hopping process
occurring just at the boundary between occupied and empty regions, leading
to an energy change of $-4U$. Under the condition $\left\vert U\right\vert
\gg \left\vert \protect\mu \right\vert, \left\vert \Delta \right\vert
,\left\vert \protect\tau \right\vert$, (c) and (d) become the predominant
processes due to the small energy gap, which ensures the validity of $H_{%
\mathrm{eff}}$ in Eq. (\ref{H_eff}).}
\label{fig1}
\end{figure}

We note that the action of $H_{\mathrm{T}}$ can couple the states within the
subspace $\left\{ \left\vert \phi _{l}\right\rangle \right\} $ to states
outside this subspace. Fig. \ref{fig1}\ schematically illustrates the energy
level structure and the transition processes between them, showing that the
effective Hamiltonian remains valid under the condition $\left\vert
U\right\vert \gg \left\vert \mu \right\vert ,\left\vert \Delta \right\vert
,\left\vert \tau \right\vert $. This conclusion will be verified numerically
in Sec. \ref{Bloch oscillations}.

\section{Quantum spin representation}

\label{Quantum spin representation}

In this section, we investigate the Hamiltonian from another perspective. We
will introduce a spin representation for the Hamiltonian in Eq. (\ref{H}) at
the symmetry point $\Delta =\tau$. This allows us to gain a clear
understanding of the previous conclusions and also provides a way to measure
the dynamics obtained in this work through a spin system in experiment. Our
strategy is as follows. While an Ising chain with transverse field is
typically investigated via the Majorana representation of the Kitaev chain,
it can also be investigated by using an alternative method: a domain-wall
excitation picture. We will show the connection between an $N$-site Ising
chain under a special boundary condition and two identical copies of the
interacting Kitaev chain on ($N-1$)-site chains. Consequently, some features
of the interacting Kitaev chain can be understood from its spin
representation.

\subsection{Mapping to a spin chain}

We start with a transverse field Ising model on a zigzag lattice. The
Hamiltonian is given by 
\begin{equation}
H_{\mathrm{spin}}=\frac{\mu }{2}\sum_{j=1}^{N-1}\sigma _{j}^{z}\sigma
_{j+1}^{z}+U\sum_{j=1}^{N-2}\sigma _{j}^{z}\sigma
_{j+2}^{z}+\tau\sum_{j=2}^{N-1}\sigma _{j}^{x}, \label{H_spin}
\end{equation}%
where $\sigma _{j}^{\alpha }$ $\left( \alpha =x,y,z\right) $ are the Pauli
operators on the $j$th site. Compared with the customary Ising model, there
is additional Ising-type coupling between next-nearest-neighbor (NNN) spins.
Unlike the three-site interactions $\sigma _{j}^{z}\sigma _{j+1}^{x}\sigma
_{j+2}^{z}$, which are still exactly solvable \cite%
{suzuki1971relationship,ZhangGang2015}, the term\ $\sigma _{j}^{z}\sigma
_{j+2}^{z}$ poses an obstacle to the exact solution in general. The
Hamiltonian is subject to open boundary conditions with vanishing transverse
field at the two end sites. This results in the conservation of the edge
spins, i.e., 
\begin{equation}
\left[ \sigma _{1}^{z},H_{\mathrm{spin}}\right] =\left[ \sigma _{N}^{z},H_{%
\mathrm{spin}}\right] =0.
\end{equation}%
Consequently, the boundary spins remain fixed in eigenstates of $\sigma
_{1}^{z}$ and $\sigma _{N}^{z}$, respectively, due to the vanishing
transverse fields at the chain ends.

The four possible configurations of the two end spins define four invariant
subspaces. We restrict our attention to the two subspaces in which the last
spin is fixed, either up or down. Each of these encompasses two of the
original four subspaces, which we index by $\rho =\uparrow ,\downarrow $
based on the state of the final spin. The basis sets of the two subspaces
take the form 
\begin{equation}
\left\{ \left\vert \phi _{j}^{\uparrow }\right\rangle \right\}
:\prod_{l=1}^{N-1}\left\vert \sigma _{l}\right\rangle _{l}\left\vert
\uparrow \right\rangle _{N},  \label{subspace1}
\end{equation}%
and%
\begin{equation}
\left\{ \left\vert \phi _{j}^{\downarrow }\right\rangle \right\}
:\prod_{l=1}^{N-1}\left\vert \sigma _{l}\right\rangle _{l}\left\vert
\downarrow \right\rangle _{N},  \label{subspace2}
\end{equation}%
respectively, where $j=1,2,...,2^{N-1}$ and $\sigma _{l}=\uparrow
,\downarrow $. Here $\left\vert \uparrow \right\rangle _{l}$ and $\left\vert
\downarrow \right\rangle _{l}$ denote the eigenstates of $\sigma _{l}^{z}$
with eigenvalues $+1$ and $-1$, respectively. It is apparent that each basis
state consists of magnetic domains in different configurations, which
themselves provide an alternative basis representation. The spin-flip
operator $p=\prod_{l=1}^{N}\sigma _{l}^{x}$ connects the two subspaces via $%
\left\{ \left\vert \phi _{j}^{\uparrow }\right\rangle \right\} =p\left\{
\left\vert \phi _{j}^{\downarrow }\right\rangle \right\} $. Owing to the
spin-flip symmetry%
\begin{equation}
\left[ p,H_{\mathrm{spin}}\right] =0.  \label{SFS}
\end{equation}

We now introduce a complete basis of domain-wall excitations. The basis
transforms as: 
\begin{eqnarray}
\left\vert \uparrow \right\rangle _{l}\left\vert \downarrow \right\rangle
_{l+1} &=&\left\vert \downarrow \right\rangle _{l}\left\vert \uparrow
\right\rangle _{l+1}=\overline{\left\vert 1\right\rangle }_{l},  \notag \\
\left\vert \uparrow \right\rangle _{l}\left\vert \uparrow \right\rangle
_{l+1} &=&\left\vert \downarrow \right\rangle _{l}\left\vert \downarrow
\right\rangle _{l+1}=\overline{\left\vert 0\right\rangle }_{l},
\label{mapping}
\end{eqnarray}%
where $\overline{\left\vert 1\right\rangle }_{l}=c_{l,\rho }^{\dag }%
\overline{\left\vert 0\right\rangle }$ and $c_{l,\rho }\overline{\left\vert
0\right\rangle }_{l}=0$, with $c_{l,\rho }^{\dag }$ ($c_{l,\rho }$)\ being
the creation (annihilation) operator of a spinful fermion. Both domain-wall
types and both magnetic-domain types map to the same fermionic and vacuum
states, respectively. Despite the apparent bijectivity suggested by the dual
mapping in Eq. (\ref{mapping}), injectivity holds only within the invariant
subspace: fixing the last spin state determines the last domain-wall type,
forcing the remaining walls to alternate and thereby fixing all domain
types. Two illustrative examples for $N=8$ are:%
\begin{eqnarray}
&&\left\vert \uparrow \right\rangle _{1}\left\vert \downarrow \right\rangle
_{2}\left\vert \downarrow \right\rangle _{3}\left\vert \uparrow
\right\rangle _{4}\left\vert \uparrow \right\rangle _{5}\left\vert \uparrow
\right\rangle _{6}\left\vert \downarrow \right\rangle _{7}\left\vert
\uparrow \right\rangle _{8}  \notag \\
&=&\overline{\left\vert 1\right\rangle }_{1}\overline{\left\vert
0\right\rangle }_{2}\overline{\left\vert 1\right\rangle }_{3}\overline{%
\left\vert 0\right\rangle }_{4}\overline{\left\vert 0\right\rangle }_{5}%
\overline{\left\vert 1\right\rangle }_{6}\overline{\left\vert 1\right\rangle 
}_{7},
\end{eqnarray}%
and%
\begin{eqnarray}
&&\left\vert \uparrow \right\rangle _{1}\left\vert \uparrow \right\rangle
_{2}\left\vert \downarrow \right\rangle _{3}\left\vert \downarrow
\right\rangle _{4}\left\vert \downarrow \right\rangle _{5}\left\vert
\uparrow \right\rangle _{6}\left\vert \downarrow \right\rangle
_{7}\left\vert \downarrow \right\rangle _{8}  \notag \\
&=&\overline{\left\vert 0\right\rangle }_{1}\overline{\left\vert
1\right\rangle }_{2}\overline{\left\vert 0\right\rangle }_{3}\overline{%
\left\vert 0\right\rangle }_{4}\overline{\left\vert 1\right\rangle }_{5}%
\overline{\left\vert 1\right\rangle }_{6}\overline{\left\vert 0\right\rangle 
}_{7},
\end{eqnarray}%
residing in the $\rho =\uparrow ,\downarrow $ subspaces, respectively.

We now translate the Ising-chain Hamiltonian into the fermionic language.
Physically, $\sigma _{j}^{x}$ corresponds to domain-wall pair creation or
hopping, and $\sigma _{j}^{z}\sigma _{j+1}^{z}$ yields the on-site energy
(chemical potential) for domain walls. Formally, this equivalence is
expressed as:%
\begin{eqnarray}
&&\sigma _{j}^{x}\mapsto \left( c_{j-1,\rho }^{\dag }-c_{j-1,\rho }\right)
\left( c_{j,\rho }^{\dag }+c_{j,\rho }\right) ,  \notag \\
&&\sigma _{j}^{z}\sigma _{j+1}^{z}\mapsto 1-2c_{j,\rho }^{\dag }c_{j,\rho },
\notag \\
&&\sigma _{j}^{z}\sigma _{j+2}^{z}\mapsto \left( 2c_{j,\rho }^{\dag
}c_{j,\rho }-1\right) \left( 2c_{j+1,\rho }^{\dag }c_{j+1,\rho }-1\right) .
\label{equivalence}
\end{eqnarray}

Building upon these mappings, the original Ising Hamiltonian decomposes into
two independent, identical Kitaev chains of length $N-1$, each given by%
\begin{eqnarray}
&&H_{\text{\textrm{eq}}}^{[\rho ]}=\tau\sum_{j=1}^{N-2}\left( c_{j,\rho
}^{\dagger }c_{j,\rho +1}+c_{j,\rho }^{\dagger }c_{j+1,\rho }^{\dagger
}\right) +\mathrm{H.c.}  \notag \\
&&+\sum_{j=1}^{N-2}U\left( 2c_{j,\rho }^{\dag }c_{j,\rho }-1\right) \left(
2c_{j+1,\rho }^{\dag }c_{j+1,\rho }-1\right)  \notag \\
&&-\mu \sum_{j=1}^{N-1}\left( c_{j,\rho }^{\dag }c_{j,\rho }-{\frac{1}{2}}%
\right) ,  \label{H_eq}
\end{eqnarray}%
with $\rho =\uparrow ,\downarrow $. We stress that this correspondence does
not constitute a Jordan-Wigner transformation, and the explicit mapping from
spin operators to spinful fermions remains unknown. Importantly, both $H_{%
\text{\textrm{eq}}}^{[\uparrow ]}$ and $H_{\text{\textrm{eq}}}^{[\downarrow
]}$ are formally identical to $H$ in Eq. (\ref{H}).

\subsection{Exact solutions}

Next, we consider the implications of the spin representation. In the case
with $\mu =0$, the Hamiltonian $H_{\mathrm{spin}}$\ describes two
independent transverse field Ising chains, which can be diagonalized exactly
by applying two Jordan-Wigner and Majorana transformations. In this sense,
the original interacting Kitaev Hamiltonian is in a topologically
non-trivial phase for $\left\vert U \right\vert<\left\vert \tau \right\vert $
and in a topologically trivial phase for $\left\vert U \right\vert
>\left\vert \tau \right\vert $. This is consistent with the main result of the previous work
and demonstrates the advantage of the spin representation \cite{Miao2017}.

\subsection{Ferromagnetic-antiferromagnetic domain walls}

Now we turn to the spin representation of states corresponding to the set of
fermion states $\left\{ \left\vert \phi _{l}\right\rangle \right\} $. This
enables us to understand the physical picture of the WSL in a spin system.
Based on the mapping given in Eq. (\ref{mapping}), we find that an
unoccupied fermion state on a chain corresponds to a ferromagnetic state,
which a fully occupied fermion state corresponds to an antiferromagnetic
state (N\'{e}el state). In this sense, the state $\left\vert \phi
_{l}\right\rangle $\ corresponds to a domain wall state. It is different
from that in previous works \cite{Ma2026dynamic}, where both domains are
ferromagnetic but with opposite directions. The main difference between the
two Wannier-Stark ladder systems is that the present model lacks a
longitudinal field.

\begin{figure*}[tbh]
\centering
\includegraphics[width=1.0\textwidth]{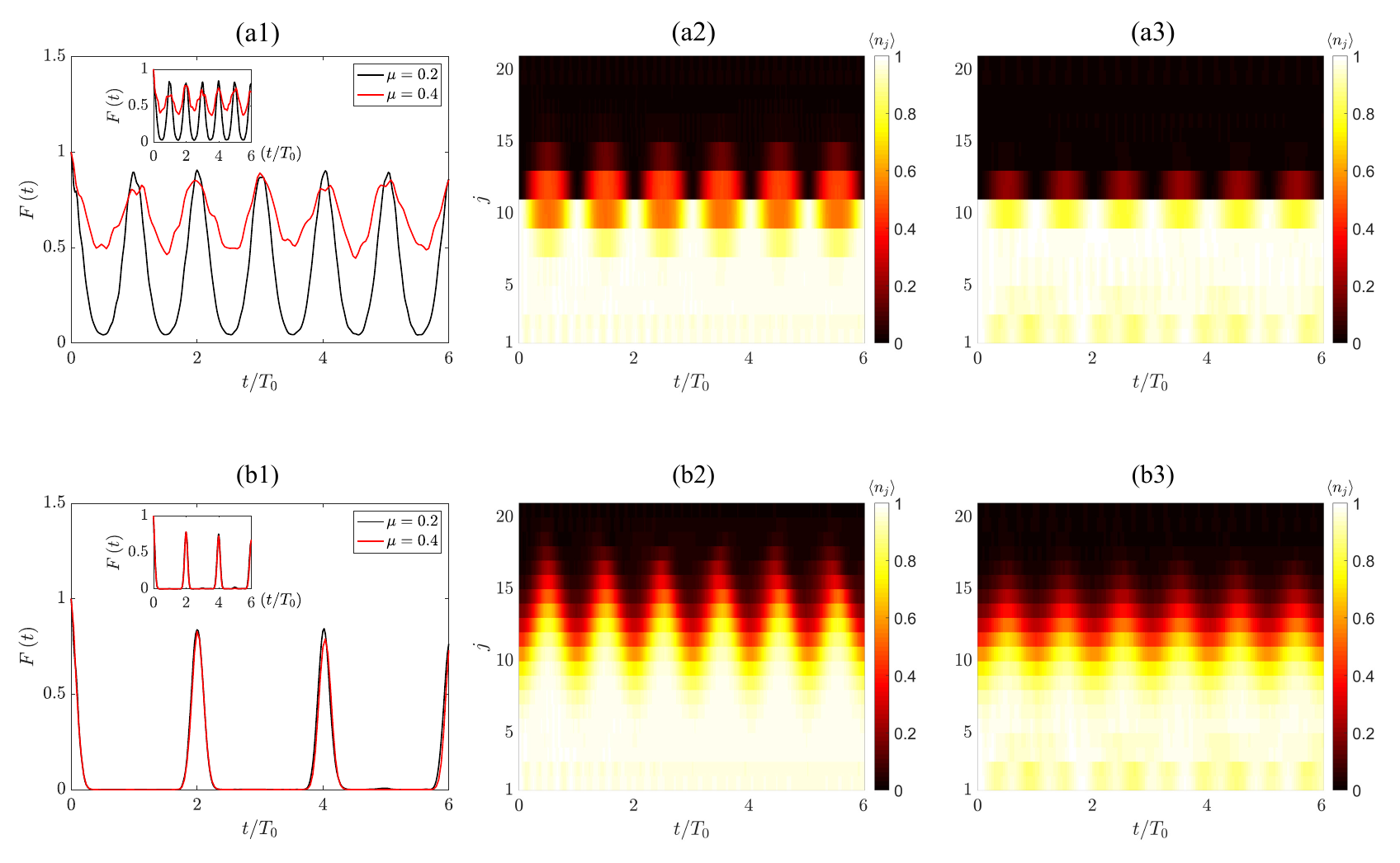}
\caption{Plots of the fidelity defined in Eq. (\protect\ref{F}) and the
average value of the particle number operator given in Eq. (\protect\ref%
{particle number}) for two kinds of initial states. Panels (a1) and (b1)
display the fidelity for the Kronecker-delta initial state and Gaussian-type
initial state with different on-site potentials, respectively, where the other
parameters for the Hamiltonian are $U=1$, $\protect\tau=0.2$, $\Delta=0.2$,
and $N=20$. For the Kronecker-delta initial state, $l_{0}=N/2$, and for the 
Gaussian-type initial state, $\alpha=5$.
The insets in (a1) and (b1) show the cases for $U=1$, $\protect%
\tau=0.8$, and $\Delta=0.2$. Here, $T_{0}=\protect\pi/\protect\mu$ is the
theoretical period of the Bloch Oscillation, and it is distinct for
different parameter settings of $\protect\mu$. Panels (a2) [(b2)] and (a3)
[b3] present the corresponding particle number of the initial state in (a1)
[b1] under the driving Hamiltonian with $\protect\tau=0.2$ while $\protect\mu%
=0.2$ and $\protect\mu=0.4$, respectively. The results in (a1) and (b1)
indicate that the dynamics are not highly sensitive to the hopping strength $%
\tau$ because, for an arbitrary initial state $\left\vert \protect\phi %
_{l}\right\rangle$, hopping can occur only at the $l$-th site and leads to
an energy difference $\Delta E=4U\gg 2\protect\mu$; therefore, only a very
large $\tau$ substantially affects the dynamics. For given parameters $U$, $%
\Delta $ and $\tau$, the Kronecker-delta initial state shows oscillatory
amplitude of the fidelity that remarkably decreases when $\protect\mu $
increases, because transitions between different levels are suppressed due
to the increased energy gap. However, the Gaussian-type initial state is a
relatively extended state, which makes it insensitive to changes in $%
\protect\mu $ to some extent. Panels (a2) [(b2)] and (a3) [b3] show that the 
average particle number periodically oscillates with period $T_{0}$ for both initial 
states, and the periodic behavior is more evident for smaller $\mu$. This is because the 
particle number operator preserves the parity of a state; therefore, 
the period is determined only by the odd or even parity part of the evolved state.}
\label{fig2}
\end{figure*}

\begin{figure*}[tbh]
\centering
\includegraphics[width=1.0\textwidth]{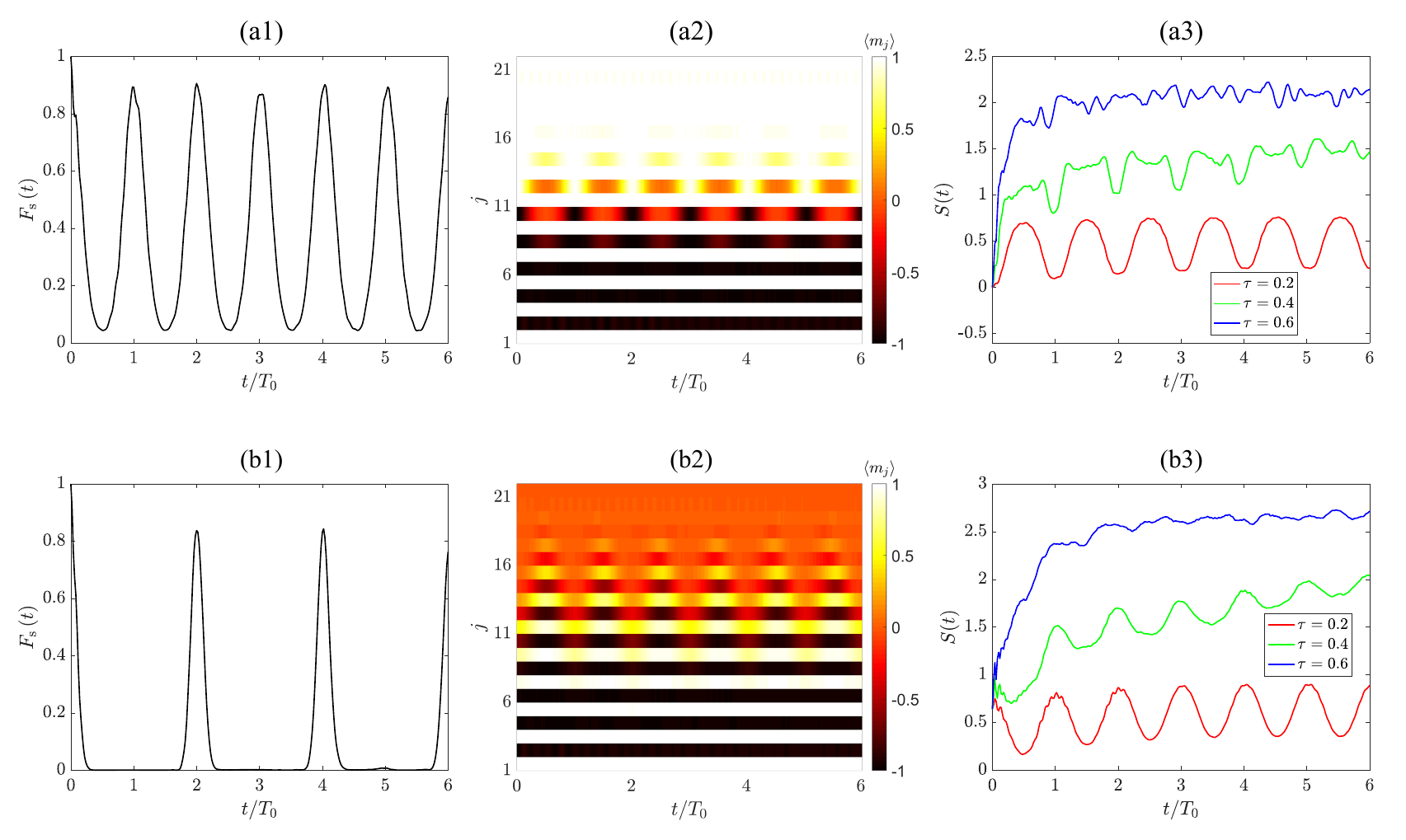}
\caption{Plots of fidelity defined in Eq. (\protect\ref{Fs}), local
magnetization given by Eq. (\protect\ref{m_i}), and entanglement entropy
given by Eq. (\protect\ref{S(t)}) in the corresponding spin model given by
Eq. (\protect\ref{H_spin}) of the Kitaev model with $\Delta=\protect\tau$.
Panels (a1), (a2), and (a3) correspond to the Kronecker-delta initial state,
and panels (b1), (b2), and (b3) correspond to the Gaussian-type initial
state. Other parameters are $U=1$, $\protect\tau=0.2$, $\protect\mu=0.2$ and 
$N=21$. These results indicate that the entanglement entropy oscillates
around a small value with period $T_{0}=\protect\pi/\protect\mu$ for small $%
\protect\tau$. However, it increases for large $\tau$ and remains at
an approximately fixed value due to the finite-$N$ effect. }
\label{fig3}
\end{figure*}

\section{Bloch oscillations}

\label{Bloch oscillations}

The periodic dynamics of WSL systems, characterized by Bloch oscillations,
are firmly established. From an experimental standpoint, spin systems offer
greater feasibility than Kitaev systems, in which particle number
conservation is violated. In each invariant subspace, the dynamics driven by
the effective Hamiltonian in Eq. (\ref{H_eff}) is governed by the kernel in
the form 
\begin{equation}
K_{jj^{\prime }}\left( t\right) =\sum_{n}e^{-iE_{n}t}\left\langle j|\psi
_{n}\right\rangle \left\langle \psi _{n}|j^{\prime }\right\rangle .
\end{equation}%
In the thermodynamic limit, straightforward derivation shows that 
\begin{eqnarray}
K_{jj^{\prime }}\left( t\right) &=&\sum_{n}e^{i2n\mu t}J_{j-n}\left( \frac{%
\Delta }{\mu }\right) J_{j^{\prime }-n}^{\ast }\left( \frac{\Delta }{\mu }%
\right)  \notag \\
&=&\left( i\right) ^{j-j^{\prime }}e^{i\left( j+j^{\prime }\right) \mu
t}J_{j-j^{\prime }}\left[ \frac{-2\Delta \sin \left( \mu t\right) }{\mu }%
\right]
\end{eqnarray}%
by applying Graf's addition theorem \cite%
{abramowitz1966handbook,holthaus1996localization}. We note that $%
K_{jj^{\prime }}\left( t\right) $ is a periodic function satisfying $%
K_{jj^{\prime }}\left( t+T_{0}\right) =K_{jj^{\prime }}\left( t\right) $,
with period $T_{0}=\pi /\mu $. This indicates that for
any initial state belonging to the subspace with even or odd particle number,
i.e., $\left\vert \psi (0)\right\rangle =\left\vert \psi _{\text{\textrm{e}}%
}(0)\right\rangle $, or $\left\vert \psi (0)\right\rangle =\left\vert \psi _{%
\text{\textrm{o}}}(0)\right\rangle $, we have $\left\vert \psi
(t+T_{0})\right\rangle \propto \left\vert \psi (t)\right\rangle $. However, for any
initial state belonging to both subspaces, i.e., $\left\vert \psi
(0)\right\rangle =\left\vert \psi _{\text{\textrm{e}}}(0)\right\rangle
+\left\vert \psi _{\text{\textrm{o}}}(0)\right\rangle $, we have $\left\vert
\psi (t+2T_{0})\right\rangle \propto \left\vert \psi (t)\right\rangle $ since the
even and odd components acquire opposite signs after one period, $\left\vert
\psi (t+T_{0})\right\rangle \propto \left\vert \psi _{\text{\textrm{e}}%
}(T_{0})\right\rangle -\left\vert \psi _{\text{\textrm{o}}%
}(T_{0})\right\rangle $.

To examine the dynamical behavior, we consider the temporal evolution for
the following two initial states $\left\vert \psi (0)\right\rangle
=\sum\nolimits_{l}d_{l}\left\vert \phi _{l}\right\rangle $:

(i) a state initially localized at site $l=l_{0}$, referred to as
Kronecker-delta state with $d_{l}=\delta _{l,l_{0}}$;

(ii) a Gaussian-type initial state with $d_{l}=\Omega ^{-1}e^{il\pi
/2}e^{-(l-N/2)^{2}/\alpha ^{2}}$, where $\Omega $\ is the normalized
coefficient.

We can estimate the result from the above analysis. We find that the delta
initial state should exhibit Bloch oscillation with period $T_{0}$, while
the Gaussian initial state exhibits it with period $2T_{0}$. For finite-size
system, we carry out numerical simulations of the temporal evolution by the
exact diagonalization.

We introduce fidelity, defined as 
\begin{equation}
F\left( t\right) =\left\vert \left\langle \psi \left( 0\right) |\psi \left(
t\right) \right\rangle \right\vert ^{2},  \label{F}
\end{equation}%
and the average value of the particle number operator, defined as%
\begin{equation}
\left\langle n_{j}\right\rangle =\left\langle \psi \left( t\right)
\right\vert c_{j}^{\dag }c_{j}\left\vert \psi \left( t\right) \right\rangle ,
\label{particle number}
\end{equation}%
to characterize Bloch oscillation. Numerical results for several
representative parameter sets are displayed in Fig. \ref{fig2},
demonstrating clear Bloch oscillations. Specifically, the oscillatory period
is $\pi/ \mu$ for the Kronecker-delta initial state and $2\pi/ \mu$ for the
Gaussian-type initial state. However, the average particle number exhibits periodic oscillation with period $T_{0}=\pi/ \mu$ for both initial states, which is determined only by the period of the odd or even particle number subspace wave function rather than the period of the whole wave function. These results are consistent with our theoretical
predictions and verify the conclusion drawn from the analysis of
the energy structure in Fig. \ref{fig1}.

In parallel, we explore the dynamical behaviors within the corresponding
spin system, which may serve as an experimental protocol for verification.%
\textbf{\ }This procedure consists of the following steps. First, the
initial state is mapped onto the spin representation using the
transformation defined in Eq. (\ref{mapping}): $\left\vert \psi \left(
0\right) \right\rangle \rightarrow \left\vert \varphi \left( 0\right)
\right\rangle$. Second, to characterize the time evolution, we calculate
the fidelity 
\begin{equation}
F_{\mathrm{s}}\left( t\right) =\left\vert \left\langle \varphi \left(
0\right) |\varphi \left( t\right) \right\rangle \right\vert ^{2},  \label{Fs}
\end{equation}%
where the time-evolved state is given by 
\begin{equation}
\left\vert \varphi \left( t\right) \right\rangle =e^{-iH_{\mathrm{spin}%
}t}\left\vert \varphi \left( 0\right) \right\rangle .
\end{equation}%
Third, we compute two observables, the magnetization

\begin{equation}
m_{j}=\left\langle \varphi (t)\right\vert \sigma _{j}^{z}\left\vert \varphi
(t)\right\rangle ,  \label{m_i}
\end{equation}%
for $j\in \left[1, N \right] $, and the bipartite Von Neumann entropy is defined by

\begin{equation}
S(t)=-\mathrm{Tr}\left( \rho _{\mathrm{A}}\ln \rho _{\mathrm{A}}\right) ,
\label{S(t)}
\end{equation}%
where%
\begin{equation}
\rho _{\mathrm{A}}=\text{\textrm{Tr}}_{\text{\textrm{B}}}\left( \left\vert
\varphi \left( t\right) \right\rangle \left\langle \varphi \left( t\right)
\right\vert \right) ,
\end{equation}
is reduced density matrix for sublattice A. In this work, A and B denote
left and right half of the chain of sublattices, respectively.

The numerical results presented in Fig. \ref{fig3} show that the fidelity
revives periodically and that the domain wall between the ferromagnetic
region and the N\'{e}el region exhibits periodic oscillations with period $T_{0}$ for 
both the Kronecker-delta and the Gaussian-type initial states. This agrees with the results
obtained for the interacting Kitaev model.

\section{Summary}

\label{Summary}

In summary, we have studied the possible impact of the NN interaction on the
dynamics of the Kitaev chain. We have shown that the interplay between
the pairing processes and the NN interaction can result in the formation of
a WSL. This demonstrates that Bloch oscillations can occur in a system
without particle number conservation but with translational symmetry. To
gain a clear physical picture and guide experimental verification, we map
the model at the symmetry point onto a transverse field Ising model on a
zigzag lattice via the spin-fermion correspondence. In this spin
representation, the Wannier-Stark state corresponds to a localized domain
wall between the ferromagnetic and antiferromagnetic phases. The underlying
mechanism of the Bloch oscillation of the domain wall is not induced by the
longitudinal field, as in previous works, but rather by the pairing process.
Numerical simulations of time-dependent observables verify these
conclusions. Our findings provide an example demonstrating emergent
Wannier-Stark\ many-body localization, where Wannier-Stark states are
localized due to intrinsic interactions rather than disorder or a strong
tilted potential.

\acknowledgments This work was supported by the National Natural Science
Foundation of China (under Grant No. 12374461).

\bibliography{Kitaev_BO_references.bib}

\end{document}